\documentclass{article}
\usepackage{amsmath,amssymb}
\usepackage[text={7in,9.5in},centering]{geometry}
\usepackage{subfigure}
\usepackage{graphicx}
\usepackage{tabularx}
\usepackage{cite}

\usepackage{color}

\newcommand{\figpath}{./}

\newlength{\figwidth}
\newcommand{\citep}[1]{{\cite{#1}}}
\newcommand{\cref}[1]{{\cite{#1}}}

\newcommand{\eref}[1]{{Eq. (\ref{#1})}}

\newcommand{\sref}[1]{{Sec. \ref{#1}}}
\newcommand{\fref}[1]{{Fig. \ref{#1}}}

\newcommand{\ie}{{\it i.e.\,}}
\newcommand{\vs}{{\it vs.\,}}
\newcommand{\eg}{{\it e.g.\ }}
\newcommand{\etal}{{\it et al.\,}}

\newcommand{\ab}{\mathbf{a}}
\newcommand{\bb}{\mathbf{b}}
\newcommand{\cb}{\mathbf{c}}

\newcommand{\eb}{\mathbf{e}}

\newcommand{\nb}{\mathbf{n}}

\renewcommand{\sb}{\mathbf{s}}

\newcommand{\ub}{\mathbf{u}}

\newcommand{\wb}{\mathbf{w}}
\newcommand{\xb}{\mathbf{x}}
\newcommand{\yb}{\mathbf{y}}
\newcommand{\zb}{\mathbf{z}}
\newcommand{\Ab}{\mathbf{A}}

\newcommand{\Eb}{\mathbf{E}}
\newcommand{\Fb}{\mathbf{F}}

\newcommand{\Ib}{\mathbf{I}}

\newcommand{\Lb}{\mathbf{L}}

\newcommand{\Nb}{\mathbf{N}}

\newcommand{\Pb}{\mathbf{P}}

\newcommand{\Sb}{\mathbf{S}}

\newcommand{\Xb}{\mathbf{X}}

\newcommand{\Cbb}{\mathbb{C}}

\newcommand{\chib}{\boldsymbol{\chi}}
\newcommand{\xib}{{\boldsymbol{\xi}}}

\newcommand{\farg}[1]{\left(#1\right)}
\newcommand{\abrkt}[1]{\left<#1\right>}

\newcommand{\sgn}{\operatorname{sgn}}

\newcommand{\shearmod}{G}
\newcommand{\stress}{{\boldsymbol{\sigma}}}

\newcommand{\distortion}{{\boldsymbol{\beta}}}

\newcommand{\config}{{\kappa}}

\newcommand{\dislocationdensity}{{\boldsymbol{\alpha}}}
\newcommand{\loc}{\Delta}

\newcommand{\dr}{\mathrm{d}}
\newcommand{\grad}{\boldsymbol{\nabla}_\mathbf{x}}
\newcommand{\Grad}{\boldsymbol{\nabla}_\mathbf{X}}

\newcommand{\Curl}{\boldsymbol{\nabla}_\mathbf{X} \times}
\newcommand{\permutation}{\boldsymbol{\epsilon}}

\begin{document}
\title {\bf Comparison of dislocation density tensor fields derived from discrete dislocation dynamics and crystal plasticity simulations of torsion  \rm }
\vspace{0.1in}
\author{ 
Reese E. JONES\footnote{ {Corresponding author}: rjones@sandia.gov},  \
Jonathan A. ZIMMERMAN, \\ 
\sl Sandia National Laboratories, Livermore, CA 94550, USA \\[0.1in]
Giacomo PO, \\
\sl University of California, Los Angeles, CA 90095, USA \\[0.1in]
\rm \normalsize}
\date{
}
\maketitle
\small
\normalsize
\begin{abstract}
The importance of accurate simulation of the plastic deformation of ductile metals to the design of structures and components is well-known. 
Many techniques exist that address the length scales relevant to deformation processes, including dislocation dynamics (DD), which models the interaction and evolution of discrete dislocation line segments, and crystal plasticity (CP), which incorporates the crystalline nature and restricted motion of dislocations into a higher scale continuous field framework. 
While these two methods are conceptually related, there have been only nominal efforts focused on the system-level material response that use DD-generated information to enhance the fidelity of plasticity models. 
To ascertain to what degree the predictions of CP are consistent with those of DD, we compare their global and microstructural response in a number of deformation modes.
After using nominally homogeneous compression and shear deformation dislocation dynamics simulations to calibrate crystal plasticity flow rule parameters, we compare not only the system-level stress-strain response of prismatic wires in torsion but also the resulting {geometrically necessary} dislocation density fields. 
To establish a connection between explicit description of dislocations and the continuum assumed with crystal plasticity simulations, we ascertain the minimum length-scale at which meaningful dislocation density fields appear.
Our results show that, for the case of torsion, the two material models can produce comparable spatial dislocation density distributions.
\end{abstract}

\protect\vspace{0.1in}
\nopagebreak[4]
\begin{description}
\item[Keywords: ] Dislocation dynamics, crystal plasticity, dislocation density tensor
\end{description}



\section{Introduction}

The importance of accurate simulation of the plastic deformation of ductile metals in the design of structures and components to performance and failure criteria is well known. 
Plasticity models 
describe the influence of elastic and inelastic deformation on stress within a body that undergoes a specific displacement/loading path. 
These models are conventionally constructed with parameters, such as elastic constants, yield stress and work hardening, fitted to experimentally measured data.

Crystal plasticity (CP) is a particular form of a plasticity model that takes into account some details of the underlying crystal structure.
In CP, characteristics of the motion of dislocations (nanometer scale line imperfections in the crystal structure of a material) influence the deformation experienced by the material. 
Early CP models connected plastic strain rate to Burgers vectors, velocity and line length density of inherent material dislocations \citep{bilby1960continuous,ortiz1982statistical,asaro1983crystal}.
More recent models cast this response in a finite deformation framework, and decompose the plastic portion of the velocity gradient into separate contributions from various families of dislocations, each associated with a specific slip system \citep{kuchnicki2006efficient,lee2010dislocation,zhao2016integrated}.

Given the nanoscale nature of the underlying defects that govern plastic deformation in metals, it is logical to consider whether simulation methods that explicitly model these defects, such as atomistic simulation and discrete dislocation dynamics (DD) see \eg \cite{mordehai2008introducing} and \cite{groh2009multiscale},
can be used to improve the fidelity of CP constitutive relations. 
The connection between dislocation interaction and plasticity was pioneered by \cite{Orowan1934}, \cite{Polanyl1934}, and \cite{taylor1934mechanism,taylor1934mechanism2}.
Since that time, a number of researchers have made connections
between dislocation mechanisms and crystal plasticity,  
 see \eg \cite{depres2006low,zhou2010discrete,chandra2015multiscale}.

A number of recent efforts have made comparisons of molecular dynamics results with continuum plasticity simulations.
\cite{Horstemeyer2003} compared atomistic simulation of FCC nickel with both CP and macroscopic internal state variable theories for the case of simple shear.
This comparison revealed qualitative similarities between the different methods, but also quantitative differences due to the presence of thermal vibrations that occur for atomistic simulation at room temperature. 
In particular, it was observed that a much narrower stress distribution arose for the finite element analyses than for that extracted from atomistics. 
This connection to thermal vibration was verified through additional simulations at low temperature ($\sim$10K). 
Efforts have also been made by 
Vitek \citep{vitek2008non,vitek2011atomic}, 
Gr{\"o}ger \etal \citep{Groger2008a,Groger2008b,Groger2008c}, and Weinberger \etal \citep{Weinberger2012,lim2013application,hale2015insights} to directly use atomistic simulation results to parameterize microscale yield response and CP models in body-centered cubic (BCC) metals. 
Vitek and coworkers examined the non-Schmidt behavior of the slip systems of group VB and VIB systems using a Finnis-Sinclair inter-atomic potential.%
\footnote{
In a material exhibiting non-Schmidt behavior, plastic deformation does not align with maximum stress resolved on slip planes.}
In their work, Gr{\"o}ger \etal used bond order potentials to model the behavior of screw dislocations in Mo and W, quantifying a stress/loading-direction dependency to the critical resolved shear stress and characterizing the resulting non-Schmid behavior.
Weinberger \etal leveraged this approach to fit parameters of a single crystal yield constitutive model to zero-temperature molecular statics simulation results. 
They then combined this parameterization with experimental data to develop a temperature and stress dependent model for the evolving plastic strain rate. 
This approach is particularly well-suited for BCC metals, where lattice friction for individual dislocations represents a significant barrier to glide.

Dislocation dynamics has also been recently employed to make comparisons between continuum plasticity and lower-level simulation methods.
\cite{wang2011atomistically} built on the insights of Vitek \etal and Gr{\"o}ger \etal  in developing a three-dimensional discrete DD model to characterize the relationship between dislocation glide behavior and macro-scale plastic slip in single crystal BCC Ta. 
For FCC metals, where interactions between dislocations and with other obstacles, \eg precipitates, strongly dictate the speed of dislocation motion and resulting rate of plastic deformation, atomistic simulation is limited in its ability to simulate a significant quantity of dislocations. 
In this regime, DD can play a useful role in providing relevant characteristics used in the fitting of CP model parameters. 
Such a connection was exploited by \cite{Giessen2003}, who employed simple boundary value problems to compare the predictions of nonlocal plasticity theories with DD results. 
\cite{Groh2009} later used both atomistic simulation and DD in a hierarchy of modeling methods, with results from atomistics used to quantify individual dislocation mobilities for use in DD simulations, and DD results to quantify parameters related to work hardening within a CP framework. 
These researchers successfully used this combined approach to predict the mechanical response of an aluminum single crystal deformed under uniaxial compressive loading along the [421] crystal direction. 
The computed strain-stress response agreed well with experimental data that was not used in model calibration. 

In addition to MD and DD, other methods such as phase field crystal methods using diffusive dynamics have also proven useful in simulating plastic behavior with near atomic detail, see, \eg \cref{berry2012defect} and \cref{wang2016developing}.

While these efforts have achieved some degree of success, they have been limited in how they use the data from the evolution of explicit dislocations to enhance the fidelity of the plasticity models considered. 
In general, only the global stress-strain response of a system containing a distribution of dislocations is used to parameterize the hardening behavior; no local information is used to connect the two models. 
In particular,  a direct comparison of the dislocation density tensor field (DDT) resulting from the evolution of the dislocation distribution in the case of DD and from the plastic deformation in the case of CP has not been considered thus far.
However, the inclusion of the DDT as an independent field or a parameter in the flow rules and yield criteria of single-crystal plasticity models has been explored, see
\cite{Gurtin2002,Gupta2007,Hansen:2013ii,Leung:2015kw,Lee:2010gr,Aoyagi:2007ky},
and can be used to model length-scale effects
\citep{Stolken1998,Evers:2004jo}, predict different dislocation microstructures
\citep{Edmiston2013}, and allows comparison with EBSD and X-ray micro-diffraction data 
\citep{Larson:2008tg,Kysar:2007vu,Kysar:2010jx,Field:2010ex,Field:2010gi}.

A comparison of the dislocation density tensor field resulting from corresponding DD and CP simulations is the primary contribution of this work.
An attendant development is the means of extracting the DDT field from both methods and the {associated length}-scale at which {one may} resolve the DD-based densities.
After a review of the basic theory and the two simulation methods in \sref{sec:theory} and \sref{sec:method}, we compare corresponding simulations resulting from both methods in \sref{sec:results}.
To make the comparison meaningful we make the two models as consistent as possible given the different formulations.
First, by using nominally \textit{homogeneous} compression and shear loading simulations, we calibrate a simple, representative flow rule for the CP model based on the DD response in \sref{sec:calibration}.
To eliminate dynamic effects we consider loading rates where both the DD and CP results are relatively insensitive to changes of rate.
Another key predicate for consistency is the appropriate scale for the comparison based on the averaging volume used to translate discrete dislocations in DD into a density field.
We hypothesize that there exists a range of volumes where a field is resolved and remains practically indistinguishable for small changes in the averaging volume. 
This notion of an {\it intermediate asymptotic scale}, refer to \cite{ulz2013estimation}, for the dislocation density field is crucial to this work.
Then, after parameter calibration and estimation of minimal asymptotic scale, 
we compare the spatially-varying DDT field, as well as the global response, resulting from DD and CP simulations (\sref{sec:comparison}). 
Such comparison is performed for  torsion of wires with square cross section. 
The rationale behind this choice is that, on the one hand, torsion induces a deformation state which is \textit{inhomogeneous} within the cross section, and therefore its plastic component can be associated with \textit{geometrically necessary} dislocation microstructures. 
On the other hand, the deformation state is homogeneous along the twist axis, therefore allowing meaningful averages along this direction for all fields extracted from DD and CP.
This comparison of inhomogeneous loading in DD is enabled by method of \cite{po2014recent} and the MODEL code \citep{model}, which treats the boundary effects of a finite system with interacting dislocations.
(Other efforts along these lines can be found in \cref{devincre2015physically}.)
We conclude with a discussion of the results and insights on how aspects of CP models can be improved using DD simulation in the final section.
To our knowledge, comparison of DD and CP response at the levels of both overall stress-strain response and microstructural fields has not been investigated before.

\section{Theory} \label{sec:theory}

The dislocation density tensor (DDT, $\dislocationdensity$) introduced by \cite{Nye1953} is a measure of the geometrically necessary%
\footnote{The geometrically necessary portion of the total dislocation density is the part ensuring compatibility of the overall deformation from reference lattice to current configuration, as opposed to the statistically stored portion.}
dislocation density in a region of a crystal. 
The DDT is a fundamental quantity connecting the individual dislocations at the micro-scale to the macroscopic elastic and plastic deformation gradient fields, which are the essential kinematic ingredients in the macroscopic phenomenological theory of plasticity, see, e.g.,  \cite{LublinerBook}. 

\begin{figure}[h]
\centering
{\includegraphics[width=\figwidth]{\figpath/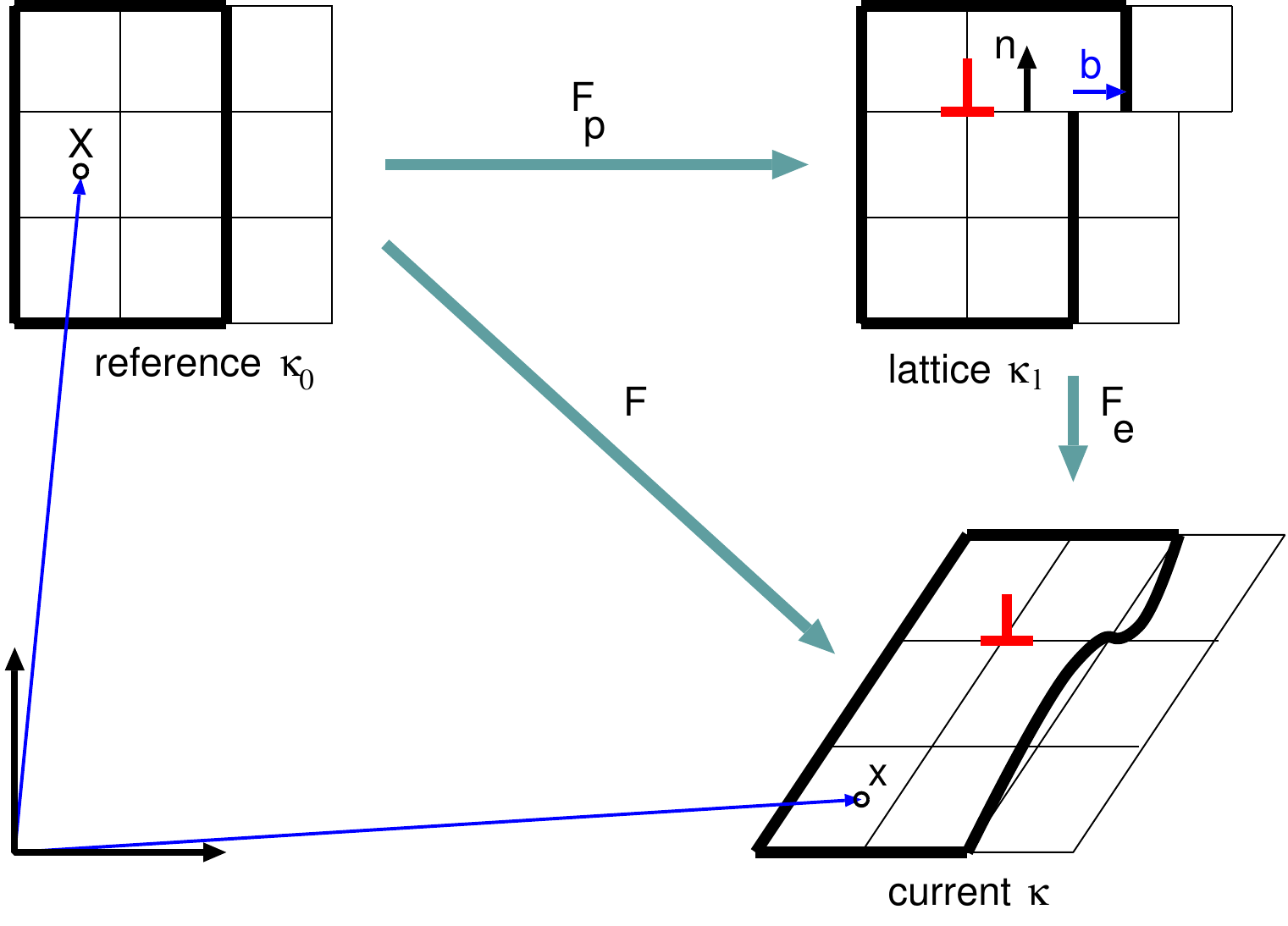}}
\caption{Reference $\kappa_0$, intermediate/lattice $\kappa_\ell$, and current $\kappa$ configurations. 
The Burgers circuit and associated area are outlined with a bold black line.
In this depiction the dislocation line tangent, $\xib$, is out of the plane, ($\xib = \partial_s \xb_\ell(s)$ with $s$ being the arc-length coordinate for line position $\xb_\ell$).
The slip plane normal $\nb$, Burgers vector $\bb$ and $\xib$ form an orthogonal triad in the lattice configuration $\kappa_\ell$.
For the edge dislocation shown the slip vector $\sb$ is aligned with $\bb$.
}
\label{fig:configurations}
\end{figure}

The dislocation density tensor $\dislocationdensity$ is defined by its relationship with the (total) Burgers vector $\bb$ for oriented area $\Ab = A \Nb$ 
\begin{equation} \label{eq:burgers_def_B}
\bb_\Nb (A)  \equiv \oint_{\partial A_{\config_{0}}} \Fb_p \, \dr\Xb 
= \int_{A_{\config_{0}}} \underbrace{\Curl \Fb_p}_{\dislocationdensity} \, \dr\Ab  
= \int_{A_{\config_\ell}}  \Curl \Fb_p \left(\Fb_p^*\right)^{-1} \, \dr\ab_\ell
\end{equation}
where $\config_0$ is the dislocation-free reference configuration and $\config_\ell$ is the intermediate (or \emph{lattice}) configuration where the loop closure $\bb$ for area $A$ is measured.%
\footnote{
The adjugate $\Fb^* = \det(\Fb) \Fb^{-T}$ is defined by Nanson's formula $\dr\ab = \Fb^* \, \dr\Ab$.
As in \cref{Cermelli2001}, the curl of a tensor field $\Fb$ is defined
$
\left(\Curl \Fb \right) \cb = \Curl \left( \Fb^T \cb \right)
$
for all constant vectors $\cb$, so that
$\Curl \Fb = \epsilon_{CBA} F_{iA,B} \Eb_C \otimes \eb_i $, where $\epsilon_{CBA}$ is the permutation tensor.
The associated version of Stokes theorem is
$
\oint \Fb \, d\Xb = \int \left(\Curl \Fb\right)^T \, \dr\Ab
$.
} 
This definition relies on Stokes theorem and the usual multiplicative decomposition, originally due to \cite{lee1969elastic}, of the deformation gradient $\Fb$,  
\begin{equation} \label{eq:FeFp}
\Fb = \Fb_e \Fb_p
\end{equation}
where the deformation gradient $\Fb = \displaystyle\Grad \chib$ is the derivative of the motion $\xb = \chib(\Xb,t)$ with respect to the reference coordinates $\Xb$,
refer to \fref{fig:configurations}.
As denoted in \eref{eq:burgers_def_B}, the DDT $\dislocationdensity$, being the curl of the plastic deformation 
\begin{equation}\label{eq:dd_macro_reduce}
\dislocationdensity 
\ = \ \Curl \Fb_p
\ ,
\end{equation}
maps (oriented) elements of area $\dr\Ab$ in the reference configuration $\kappa_0$ to infinitesimal Burgers vectors $\dr\bb$ in the incompatible lattice configuration.

The primary kinematic assumption of crystal plasticity is the rate of plastic deformation is of the form:
\begin{equation} \label{eq:plastic_rate} 
\Lb_p = \sum_a \dot{\gamma}_a \Pb_a
\Longleftrightarrow
\dot{\Fb}_p = \left(\sum_a \dot{\gamma}_a \Pb_a \right) \Fb_p \ ,
\end{equation}
where $\dot{\gamma}_a$ is the slip rate on plane $a$ and the Schmid tensors $\Pb_a = \sb_a \otimes \nb_a$ reside in the intermediate configuration and are associated with the crystallographic planes in which glissile dislocations reside.
(Here, $\sb_a$ is the slip direction and $\nb_a$ is the slip plane normal.)
This leads to an overall deformation rate given by
\begin{equation}
\Lb = \dot{\Fb} \Fb^{-1} = \Lb_e + \Fb_e \Lb_p \Fb_e^{-1}
\end{equation}
in the finite deformation context.
\fref{fig:configurations} shows one such slip plane for an edge dislocation with the slip direction $\sb$ aligned with the Burgers vector $\bb$.
In truth, \fref{fig:configurations} is a very much simplified picture of plasticity with the one dislocation shown representing a family of parallel dislocation lines as Nye envisaged; however, the essential irreversibility of macro-plasticity is a manifestation of both dislocation motion and entanglement.

Under the assumption of linearized kinematics ($\xb \approx \Xb$, small strain and rotation), the total deformation gradient can be  written as
\begin{equation} \label{eq:FeFplinear}
\Fb 
\approx \Ib+ \distortion_e+ \distortion_p
\end{equation}
where $\distortion_e=\Fb_e-\Ib$ and $\distortion_p=\Fb_p-\Ib$ are the elastic and plastic distortions, respectively. 
In particular, the process of imposing a relative displacement $\bb_\ell(\xb)$ to the two sides of a cut in surface $A_\ell$  (refer to \fref{fig:configurations}) results in the plastic distortion tensor, as in \cite{Mura:1987wt}
\begin{equation}
\distortion_p(\zb) = - \int_{A_\ell} \delta(\zb - \xb) \bb_\ell(\xb) \otimes \dr\Ab(\xb)\, .
\end{equation}
If the relative displacement is  constant over $A_\ell$, the defect is a Volterra dislocation and it models a crystal dislocation with Burgers vector $\bb_\ell$. 
In this case, the dislocation density tensor is concentrated on the closed line bounding $A_\ell$:
\begin{equation}\label{alpha_dd}
\check\dislocationdensity(\zb)  
= \grad\times\distortion_p= \oint_{\partial A_\ell} \delta(\zb - \xb) \bb_\ell \otimes \dr\xb\, ,
\end{equation}
where $\grad\times\Fb_p = \grad\times\distortion_p$.

Particularly germane to the present study is volume average of the dislocation density tensor \eqref{alpha_dd}. The average of \eref{alpha_dd} over a region $\Omega$ with volume $V$ centered at $\yb$ can  easily be obtained by virtue of the sifting property of the Dirac-$\delta$, that is
\begin{equation}\label{eq:alpha_AP}
\dislocationdensity(\yb) = \frac{1}{V}\int_{\Omega(\yb)}\check\dislocationdensity(\zb)\, \dr V(\zb)
=\frac{1}{V} \sum_{\ell\in \Omega} \bb_\ell \otimes \int_\ell 
\, \dr\xb \ .
\end{equation}
where the line integral now extends over the portion of the dislocation line inside $\Omega$.
Here, $\check\dislocationdensity(\zb)$ is at the scale of dislocations embedded in lattice and the field $\dislocationdensity(\yb)$ is at a larger length scale where continuous fields can be obtained. 
This definition classifies dislocation lines into geometrically necessary or statistically stored depending on whether the line is open or closed within the volume $\Omega$.
In previous studies \citep{Mandadapu2013} we derived a generalization of \eref{eq:alpha_AP},
\begin{equation} \label{eq:hardy_alpha}
\begin{aligned}
\dislocationdensity\farg{\yb} 
&= \ \int_\kappa \dr^3x \,  \loc\farg{\xb - \yb} \underbrace{\left[ \sum_{\ell} \bb_\ell \otimes \int_{\ell} \dr \zb \delta(\xb - \zb) \right]}_{\check\dislocationdensity(\xb)}
=   \sum_{\ell} \bb_\ell \otimes \int_\ell  \, \loc\farg{\xb - \yb}  \, \dr \xb
\\
&\approx \sum_{\ell}  \bb_\ell \otimes \Biggl[ \sum_{(ab) \in \ell} \xb_{ab}  \underbrace{\int_{0}^1 \loc\farg{\lambda \xb_{ab} + \xb_b - \yb} \dr\lambda}_{\Phi_{ab}(\yb)}  \Biggr] 
= \sum_{\ell}  \bb_\ell \otimes \left[ \sum_{(ab) \in \ell} { \xb_{ab} \Phi_{ab}\farg{\yb} } \right], 
\end{aligned}
\end{equation}
which uses bell-shaped smoothing kernels  $\loc\farg{\xb - \yb}$ that are not necessarily piecewise constant. 
This definition is evaluated using a quadrature based on straight segments $\xb_{ab} = \xb_a - \xb_b$ of length  $\ell_{ab} = \left| \frac{\mathrm{d}s}{\mathrm{d}\lambda}\right|$ with tangents $\xib_{ab} = \xb_{ab} / \ell_{ab}$. 
In a similar fashion the local line density $\rho(\yb)$ can be estimated: 
\begin{equation} \label{eq:hardy_rho}
\rho\farg{\yb} 
=  \sum_{\ell} \int_\ell  \, \loc\farg{\xb - \yb}  \, \|  \dr \xb \|
\approx 
\sum_{\ell}  \sum_{(ab) \in \ell} { \| \xb_{ab} \| \Phi_{ab}\farg{\yb} } \, .
\end{equation}
Given that the kernel function has compact support, we can use its radius of influence to estimate the range of length scales for the estimated dislocation density to be smooth and yet resolved the variations in the system.
In fact, we will estimate the minimum of the range of intermediate asymptotic scales in order to maximize the resolution of trends in the dislocation density tensor field.
In averaging via \eref{eq:hardy_alpha} to find the minimum number of dislocation lines in a given volume to obtain smooth density field and to compare with CP results, we are also making the conjecture that this is the scale of CP.

It bears mentioning that in our previous work \citep{Mandadapu2013} we demonstrated that the definition of $\dislocationdensity$, \eref{eq:alpha_AP}, satisfies the fundamental extensive property of the Burgers vector, namely that the resultant Burgers vector of a Burgers circuit is the sum of Burgers vectors of all individual dislocation lines crossing the enclosed surface. 
We also showed that Nye's original definition of dislocation density and \eref{eq:alpha_AP} become equivalent when certain conditions regarding the spacing between the dislocations and curvature of dislocations are obeyed.
Consistency, for example 
\begin{align}
\int \rho(\yb) \, \mathrm{d}V &\approx \sum_I \rho_I w_I 
\approx \sum_I \sum_{\ell}  \sum_{(ab) \in \ell} { \| \xb_{ab} \| \Phi_{ab}\farg{\yb_I} } w_I \nonumber \\
&= \sum_{\ell}  \sum_{(ab) \in \ell} { \| \xb_{ab} \| \sum_I \Phi_{ab}\farg{\yb_I} } w_I \\
&= \sum_{\ell}  \sum_{(ab) \in \ell}  \| \xb_{ab} \|  \nonumber
= \bar{\rho} V \, ,
\end{align}
where $\bar{\rho}$ is the average density, implies the requirement that
\begin{align}
\sum_I \Phi_{ab}\farg{\yb_I} w_I 
&= \sum_I \int_{0}^1 \loc\farg{\lambda \xb_{ab} + \xb_b - \yb_I} \dr\lambda \, w_I \nonumber \\
&= \int_{0}^1 \sum_I \frac{1}{V_I} N_I (\lambda \xb_{ab} + \xb_b - \yb_I) \dr\lambda \, w_I \\
&= \int_{0}^1 \sum_I N_I (\lambda \xb_{ab} + \xb_b - \yb_I) \dr\lambda 
= \int_{0}^1 \dr\lambda = 1 \nonumber
\end{align}
given (a) the localization $\loc(\yb_I) = \frac{1}{V_I} N_I$  is
based on a partition of unity $N_I(\xb) \,|\, \sum_I N_I(\xb) = 1 \ \forall \, \xb$, (b) exact integration of the line integral, and (c) the integration weights equal to kernel volumes $w_I = V_I$. 
For a piece-wise constant basis $N_I$ with compact support on a cubical region, exact integration is simply
\begin{equation}
\int_{0}^1  N_I (\lambda \xb_{ab} + \xb_b - \yb_I) \dr\lambda 
= \int_{\lambda_0}^{\lambda_1}   \dr\lambda  = \lambda_1 - \lambda_0
\end{equation}
given the two intersection points $\xb_{0|1} = \lambda_{0|1} \xb_{ab} + \xb_b$ of the segment with the boundary of the cube where $\lambda_0 < \lambda_1 \in [0,1]$. 
If either end of the segment does not intersect the cube and instead is in the interior of the cube then $\lambda_{0|1} = 0 | 1$.

\section{Method}\label{sec:method}
We employed the discrete dislocation dynamics (DD) method developed by \cite{po2014recent}, as implemented in the computer program MODEL \citep{model}. 
A brief explanation of the method is given here; a full account of the method can be found in \cref{po2014recent}.
The crystal plasticity model was implemented in the finite deformation code Albany \citep{albany}.

Given the aim of this work to compare and correlate DD and CP, we have maximized the commonalities of the two modeling techniques.
For both models, we use:
(a) isotropic elasticity with the slip planes of a cubic lattice,%
\footnote{
Isotropic linear elasticity allows use of Green's functions.
}
(b) acceleration is ignored in the balance of linear momentum, however there are dynamic/viscous effects in each model,
(c) the momentum balance is solved on finite element meshes with corresponding boundary conditions.
In this mode, both the response of CP, governed by its flow rule, and that of DD, governed by its mobility equation, are relatively insensitive to loading rate.

The formula \eqref{eq:hardy_alpha} allows to construct a continuum dislocation density field from DD simulations, which can be directly compared to the corresponding quantity defined in \sref{sec:cp} in the CP framework.  

\subsection{Dislocation dynamics}\label{sec:dd}

In general, DD approaches treat dislocation lines explicitly, with lines decomposed into discrete edges (segments) bounded by vertices. 
Given the formulas in \sref{sec:theory}, the dislocation density field $\boldsymbol\alpha(\xb)$ can be computed for a network of dislocations; however, their evolution from a given initial configuration must be computed from a closure equation yielding the dislocation velocity {$\wb$} as a function of the local stress state. 
Such an equation of motion for the discrete dislocation configuration can be obtained from thermodynamic considerations \citep{po2014recent}. 
{For FCC metals and by neglecting inertia terms, the equation of motion can be written as:} 
\begin{equation} \label{eq:PMEP}
-B\wb+\left( \stress \bb \right) \times \xib=\mathbf{0}\, .
\end{equation}
\eref{eq:PMEP} expresses the balance of the total configurational force per unit line of dislocation, and is comprised of: a viscous drag contribution,%
\footnote{
In general, the drag coefficient $B$ is a function of velocity itself, \ie the relation between drag force and dislocation velocity is non-linear; however, for the simulations presented here we assume that $B$ is a constant material coefficient. 
Also of note is the fact that the drag coefficient is in general a second-order tensor {exhibiting anisotropy depending on the crystal structure, and that  it may also depend on the local dislocation character (\ie edge versus screw), and on the type of dislocation motion (\ie glide versus climb)}. 
As described in \cref{po2014recent}, the method used here utilizes Lagrange multipliers to constrain climb processes, which are often only active at very high temperatures. Approaches that do incorporate climb include the works by \cite{gao2011investigations}, \cite{haghighat2013effect}  and \cite{babu2013dislocation}.}  
$-B\wb$, and a mechanical contribution, $\left( \stress \bb \right) \times \xib$, the Peach-Koehler force. 
In contrast to CP, the assumption of linearized kinematics adopted in DD allows to write the (small-strain) stress field $\stress$ appearing in the Peach-Koehler force as the sum of two terms:
(a) a dislocation-dislocation interaction term which can be computed accurately  using the Peach-Koehler stress equation \citep{Peach:1950tm} 
and (b) a correction term which accounts for the mechanical boundary conditions imposed to the simulation domain.

The DD formulation used in this work solves \eref{eq:PMEP} in weak/variational form  for the dislocation velocity field  using the finite element method, refer to \cite{po2014recent}. 
The numerical implementation is based on discretization of dislocation lines into segments connecting pairs of nodes. 
Segment shape functions and nodal degrees of freedom are then used for the  unknown dislocation velocity field $\wb$, so that \eref{eq:PMEP}, upon assembly over the dislocation network, is transformed into a discrete system of equations for the nodal velocities. 
Finally, the nodal velocities are used to update the nodal positions and evolve the dislocation configuration in time.

\subsection{Crystal plasticity}\label{sec:cp}

In our CP model we use Kirchhoff-St.\,Venant elasticity $\Sb = \Cbb \Eb_e$  with elastic Lagrange strain $\Eb_e = \frac{1}{2} \left( \Fb_e^T \Fb_e - \Ib \right) $.
This elastic constitutive relation reduces to 
$\stress = \Cbb \boldsymbol{\varepsilon}$ 
in the limit of small strain and rotation.
The elastic modulus $\Cbb$ with cubic symmetry is
\begin{equation}
\Cbb 
= \sum_{i,j} C_{11} \eb_{ii} \otimes \eb_{ii} 
+ C_{12} \eb_{ii} \otimes \eb_{jj} 
+ C_{44} \eb_{ij} \otimes \eb_{ij} \ ,
\end{equation}
where $\eb_{ij}$ is the dyad $\eb_i \otimes \eb_j$, which we reduce to isotropy by setting $C_{44} = \shearmod$, $C_{11} = 2 \shearmod \frac{1-\nu}{1-2\nu}$, $C_{12} = 2 \shearmod \frac{2\nu}{1-2\nu}$ for correspondence with the DD simulations.

For the plastic behavior, the kinematic assumption in \eref{eq:plastic_rate} can be seen as a plastic flow rule of the form found in \cite[Ex. 2.6]{Ortiz1999var}. 
A particularly simple hardening rule for the rate of slip $\dot{\gamma}_a$ for slip system $a$ is
\begin{equation} \label{eq:flow_rule} 
\dot{\gamma}_a = \dot{\gamma}_a (\tau_a) 
= C_a \left| \frac{\tau_a}{\tau_a^\text{crit}+H_a \gamma}\right|^{1/m} \sgn(\tau_a)
\end{equation}
subject to a $\tau_a^\text{crit}$ threshold and (isotropic) hardening modulus $H_a$.
Here $\tau_a = \Pb_a \cdot \stress$ is the shear stress resolved on the slip plane with normal $\nb_a$ and the Cauchy stress is simply
$
\stress = \frac{1}{\det \Fb} \Fb \Sb(\Eb_e) \Fb^T
$. 
The implicit update algorithm largely follows \cite{Ortiz1999var}.

Given a solution to momentum balance, we recover a dislocation density field by first using a local, element-wise L$_2$ projection to move $\Fb_p$ data at the integration points to the element nodes.
Then using the finite element basis functions $N_I$ to interpolate $\Fb_p$ on \emph{each element}, which allows differentiation with respect the reference coordinates $\Xb$, we obtain:
\begin{equation}
\dislocationdensity = \Curl \Fb_p = \permutation \cdot \Grad \Fb_p
= \sum_I \permutation  \Fb_I \Grad N_I
= \sum_{I,a,A,B,C} \epsilon_{ABC} \Fb_{(aA)I} N_{I,B} \, \eb_a \otimes \Eb_C
\end{equation}
where $I$ indexes nodes, $\permutation$ is the permutation tensor, $\eb$ and $\Eb$ are basis vectors in the spatial and reference frames (respectively), and we have dropped the subscript $p$ for clarity. 
Chain rule provides the spatial gradient $N_{I,B}$ in terms of the partial derivatives of the basis $N_I$ with respect to the local coordinates $\zeta_a$ 
\begin{equation}
N_{I,B} 
= \sum_{a} N_{I,a} \left( \partial_{\boldsymbol{\zeta}} \Xb\right)^{-1}_{aB}
= \sum_a N_{I,a} \left( \sum_J \Xb_J \otimes \partial_{\boldsymbol{\zeta}} N_J \right)^{-1}_{aB}
\end{equation}
where $\partial_{\boldsymbol{\zeta}} \Xb$ is the Jacobian of the local $\boldsymbol{\zeta}$-to-referential $\Xb$ coordinate map.
Lastly, we average the local node values to obtain a unique value for every node
and hence a continuous plastic deformation curl field.
Alternately a global L$_2$ projection could be used.

\section{Results}\label{sec:results}
For our comparison studies we chose face-centered cubic single crystals of Cu.
Its 12 slip systems are defined by the Schmid tensors $\{ \mathbf{s}_a \otimes \nb_a \} =  \{ \langle 110 \rangle \otimes \{111\}  \}$.
As mentioned, we employ an isotropic elasticity model with effective Poisson ratio $\nu = 0.34$ and  shear modulus $\shearmod = 48$ GPa.
Given that MODEL is a dimensionless code, we adopt the same normalization in our results.
Lengths are normalized to the material's Burgers vector $b = \left|\bb\right|$~=~2.556~\AA, stress-like quantities are normalized by $\shearmod$, the drag coefficient is set at $B=10^{-4}$~Pa-s, and time is normalized by $\tau = \frac{B}{\shearmod} = 2.08$ fs.

After preliminary studies, over a wide range initial (fully gissile) dislocation densities and loading rates, we chose an initial dislocation density of $\rho_0 = 10^{13}$ \text{m}$^{-2} = 6.533\times 10^{-7} \, b^{-2}$, consisting of prismatic  dislocation loops with $\abrkt{110}$-type Burgers vectors lying in $\{111\}$ planes,
 and loading rate of {$10^{-11}$} dimensionless inverse-time units (corresponding to {$4.8 \times 10^3$~sec$^{-1}$}) to approximate the rate-independent loading regime. 
From the initial dislocation density we employ a length-scale $\ell = \frac{1}{\sqrt{\rho_0}} = 1.29\times 10^3 \, b $ to non-dimensionalize $\dislocationdensity$ fields.
Given that initially the dislocation network consists entirely of loops, the total dislocation density tensor starts at zero.%
\footnote{
To allow potentially unphysical arrangement of dislocations to relax, we equilibrated the systems for 10,000 steps under zero applied load.
Although minor fluctuations in the loop geometries did occur during this simulated time, in general no major changes in the configurations were observed.
}
Note, \eref{eq:PMEP} introduces a time-scale to DD and \eref{eq:flow_rule} introduces time-scale to CP, yet neither has inertia so scaling time can lead to self-similar solutions dependent only on loading and not the rate of loading.
For DD, this limit is achieved when the dislocation interactions and collisions are sufficiently resolved.
For CP, time-scale and loading rate invariance is apparent only as $m\to\infty$ and also for small strains with a linear flow rule, as in \cite{steinmann1996numerical} and \cite{miehe2001comparative}. 
We examine the response of DD and CP for strains up to 1\%, well below the limit of validity of the linear elastic model in DD.

In \sref{sec:calibration} we use homogeneous compression and shear loading results from DD to calibrate the CP flow rule and to estimate a minimum asymptotic scale. 
Then, in \sref{sec:comparison} we compare the dislocation density fields and global response resulting from inhomogeneous loading due to torsion.
For all comparisons an ensemble of 10 replicas of the DD systems are used to alleviate sensitivity to initial dislocation arrangement.


\subsection{Calibration with homogeneous loading simulations} 
\label{sec:calibration}

With the given elastic constants and symmetries, we use compression and (simple) shear loading DD and CP simulations to calibrate the hardening parameters of the flow rule specified in \eref{eq:flow_rule}.
For these nominally homogeneous loading cases we employed a cube of dimensions $2000b\times2000b\times2000b$ oriented with the $\langle 100 \rangle$ crystal directions.
In compression, the top face of the cube (the face with outward normal in the $+\eb_3$ direction) is displaced in $\eb_3$ and the bottom face $x_3$ (face with normal in the $-\eb_3$ direction) is fixed;
the $+x_1$ and $+x_2$ lateral faces are constrained in their normal directions 
to promote homogeneous deformation and allow for expansion.
In shear, the top face is displaced in $x_1$ while fixed in $x_2$ and $x_3$;
the bottom face is fixed in $x_1$, $x_2$ and $x_3$ and all lateral faces are fixed in $x_2$ and $x_3$.

\begin{figure}[h!]
\centering
{\includegraphics[width=\figwidth]{\figpath/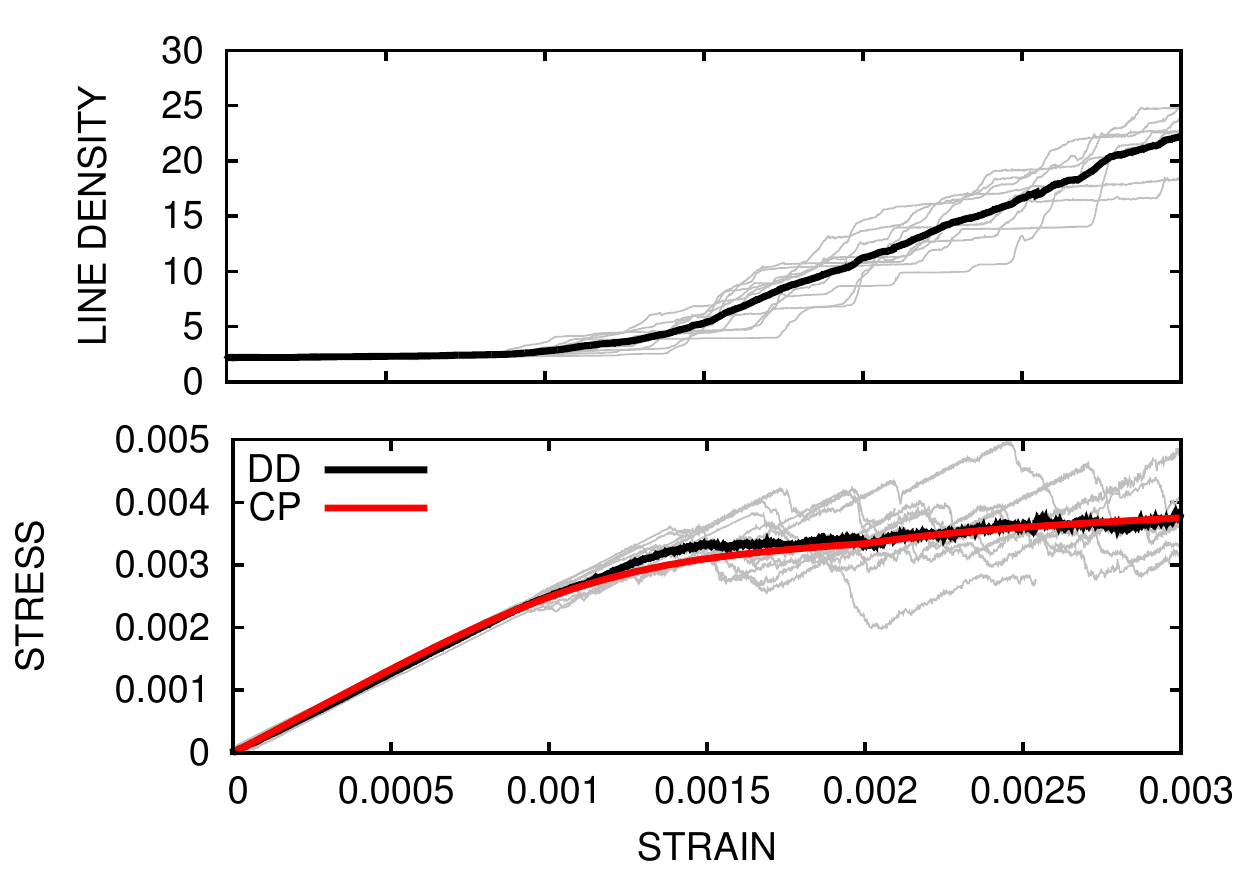}}
\caption{{Compression:} (top) dislocation line density and (bottom) 11-stress versus 11-strain.
Note the response of the individual DD simulations are shown in gray, line density is normalized by $\rho_0$ and stress by $\shearmod$.
}
\label{fig:compression_reaction_vs_strain}
\end{figure}

\begin{figure}[h!]
\centering
{\includegraphics[width=\figwidth]{\figpath/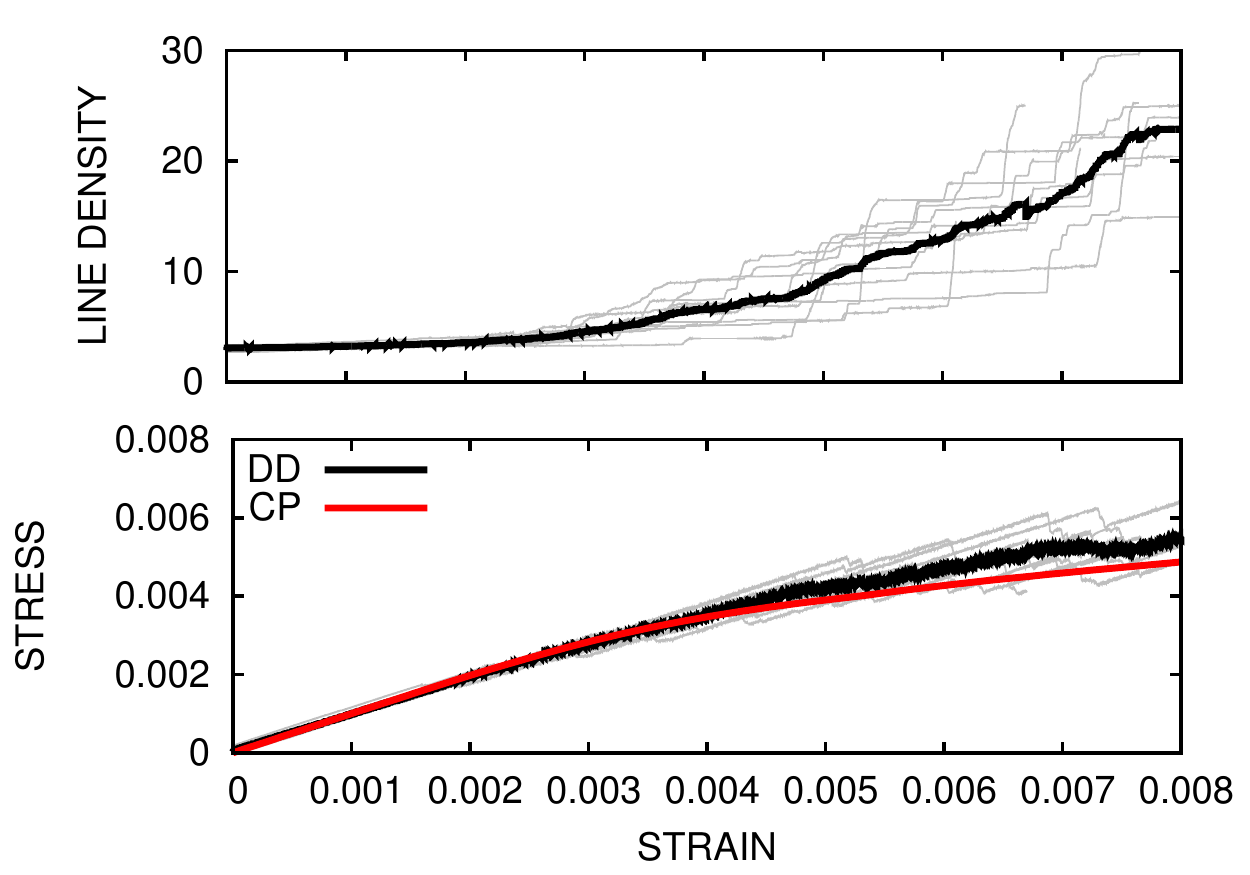}}
\caption{Shear: (top) dislocation line density and (bottom) 12-stress versus 12-strain .
Note the response of the individual DD simulations are shown in gray, line density is normalized by $\rho_0$ and stress by $\shearmod$.
}
\label{fig:shear_reaction_vs_strain}
\end{figure}

\fref{fig:compression_reaction_vs_strain} shows the stress-strain response for compression.
The initial slope of the DD response, 2.54$\shearmod$, compares well with the expected Young's modulus of $2(1+\nu)\shearmod$= 2.68$\shearmod$. 
Likewise, \fref{fig:shear_reaction_vs_strain} shows the stress-strain response for simple shear loading. 
In this case, the initial slope is 0.92$\shearmod$ near the  expected value 1.00$\shearmod$.
These results imply that the initial distribution of dislocations minimally affects the material's elastic response.  
After approximately 0.001 strain in compression and 0.004 strain in shear the total line density starts to rise rapidly which corresponds to a distinct decrease in the slope of the stress-strain response.
Using direct fits to the DD response we determined that the slope of this plastic regime was 0.12$\shearmod$ in the compression case and 0.42$\shearmod$ in the shear case and that the onset of plastic response was at 0.00125 and 0.0031 strain, respectively.
These values were used as initial guesses in calibrating the CP model.
\fref{fig:compression_reaction_vs_strain} and \ref{fig:shear_reaction_vs_strain} also show that the calibrated  CP stress-strain response compares well to that of DD using the hardening parameters: $C_a$ =1 (inverse time units), $\tau_a^\text{crit}$ = 0.11 GPa  = 0.0022 $\shearmod$, and $H_a$= 30.5 GPa  = 0.635 $\shearmod$, for all slip systems $a \in [1,12]$.
We employed the exponent $m$=4 which was the highest value that offered numerical stability.
(The issue of instability upon approaching the rate independent limit $m\to\infty$ is well-known, as discussed by \cite{steinmann1996numerical}, \cite{miehe2001comparative} and \cite{Borja2013}.)


\begin{figure}[htpb]
\centering
{\includegraphics[width=\figwidth]{\figpath/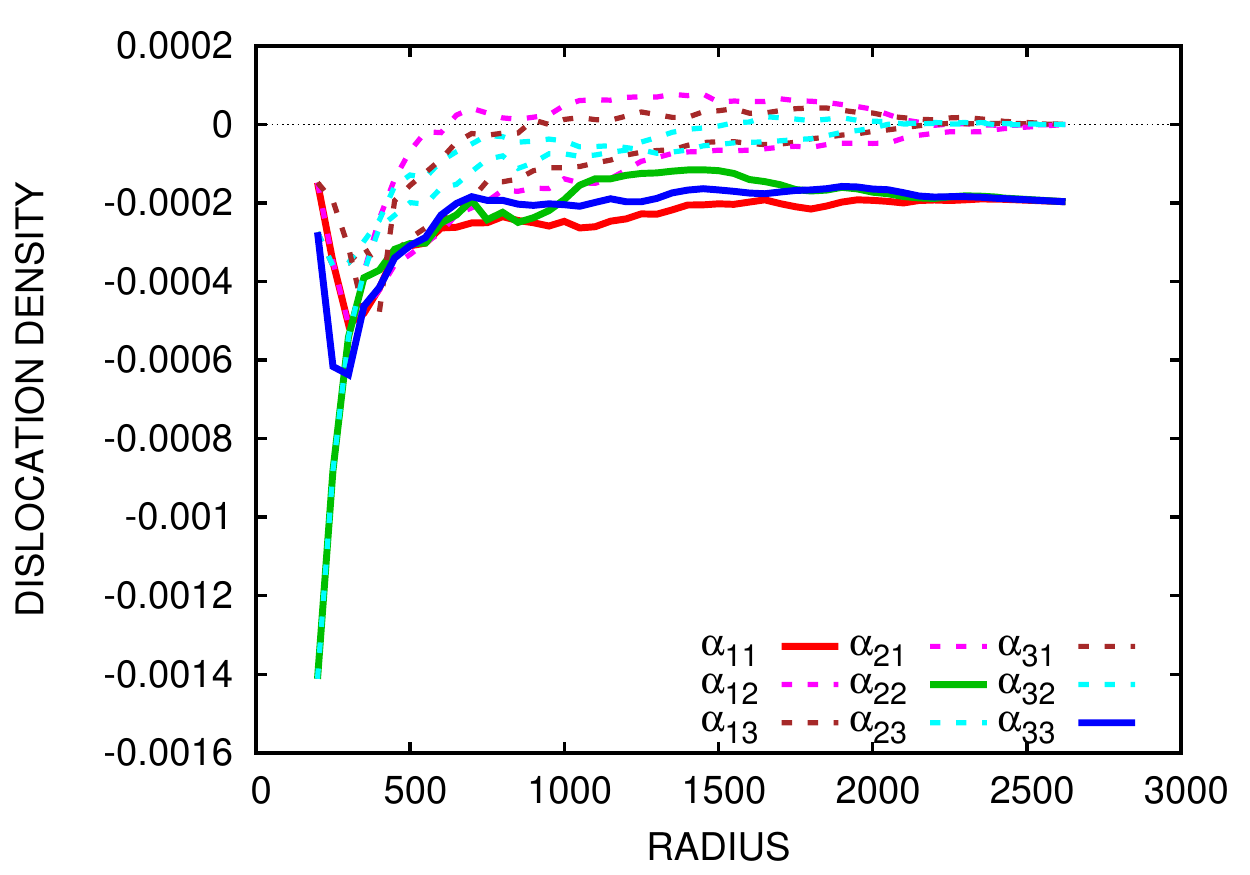}}
\caption{
Dislocation density tensor components \vs kernel radius for 3.25 percent compressive strain at an interior point in the system.
The apparent minimum asymptotic scale is approximately 500$b$ (1/8 of the system size).
Note that the values of $\dislocationdensity$ are non-dimensionalized by $\ell = 1/\sqrt{\rho}$.
}
\label{fig:asymptotic_scale}
\end{figure}

Lastly, by using a kernel estimator, \eref{eq:hardy_alpha}, positioned near the center of a $4000b\times4000b\times4000b$ system we were able to estimate the asymptotic region for the dislocation density tensor field using the compression case. 
\fref{fig:asymptotic_scale} shows that three relevant components of $\dislocationdensity$  converge to near constant, non-trivial values for kernel radii above $\approx$ 500 $b$, which is also well below the system size used for this study. 

\subsection{Comparison of inhomogeneous loading simulations} 
\label{sec:comparison}

For comparing the predictions of the DD and CP models we simulated the torsion of a $L_1\times L_2\times L_3$ = $2000b\times2000b\times4000b$ rectangular wire. 
The torsion was effected by a twisting displacement of the top relative to the bottom faces, with the bottom face fully fixed and the lateral faces traction-free.
Both $\langle 100 \rangle$ and $\langle 110 \rangle$ crystallographic orientations of the lateral faces of the wire are explored. 
Since the elastic response is isotropic, only slip plane orientation relative to the loading distinguishes these two cases.
The general solution for the stress and displacement fields in an isotropic elastic wire due to torsion are: 
\begin{equation}
\stress = \frac{1}{2}  \shearmod \theta \left(
(w,_1 -x_2) (\eb_1 \otimes \eb_3 + \eb_3 \otimes \eb_1) + 
(w,_2 +x_1) (\eb_2 \otimes \eb_3 + \eb_3 \otimes \eb_2) \right)
\end{equation}
and 
\begin{equation}
\ub = - \theta x_3 ( x_2 \eb_1 - x_1 \eb_2) + \theta w \eb_3
\end{equation}
in terms of a warping function $w=w(x_1,x_2)$.
Given the warping function characteristic of our prismatic wire with square cross-section, the torsional modulus $K$ relating the torsional moment $M \eb_3 = \int \xb \times \stress \, \mathrm{d}\mathbf{A}$ to the twist per axial distance $\theta/L_3$ is
\begin{equation}
K =  \shearmod L_1^4 \left( \frac{1}{3} - \frac{64}{\pi^3} \sum_{n=0}^\infty \frac{\tanh \pi (2n+1)/2}{(2n+1)^5} \right) \approx 0.1406 L_1^4 \shearmod
\end{equation}
where $L_1=L_2$ is the side length of the square cross-section.

\begin{figure}[htbp]
\centering
{\includegraphics[width=\figwidth]{\figpath/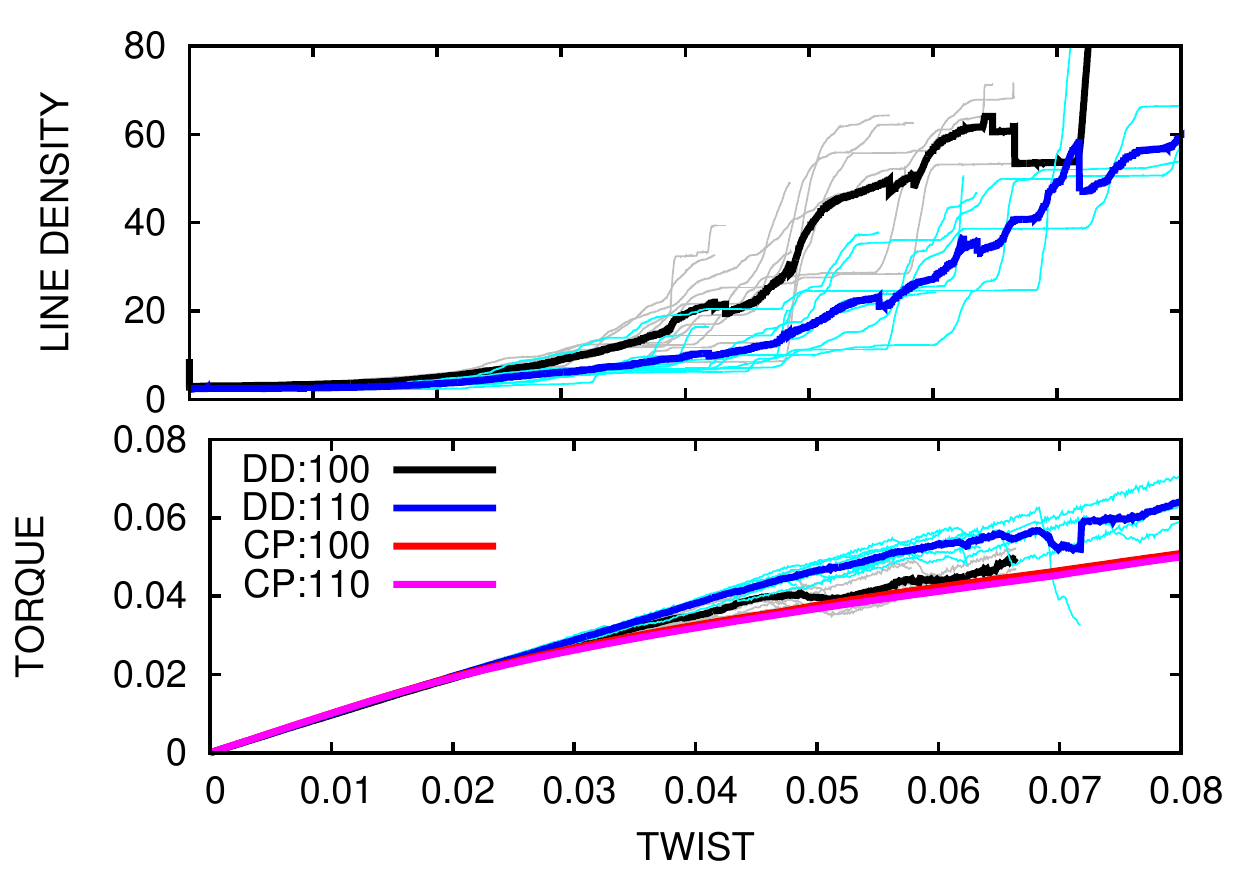}}
\caption{{Torsion:} (top) normalized dislocation line length normalized by the line density $\rho$, and (bottom) torque normalized by the torsion modulus $K$ versus twist angle (in radians).}
\label{fig:torsion_reaction}
\end{figure}

\begin{figure}[htbp]
\centering
\subfigure[$\theta = 0.0$]
{\includegraphics[width=0.40\textwidth]{\figpath/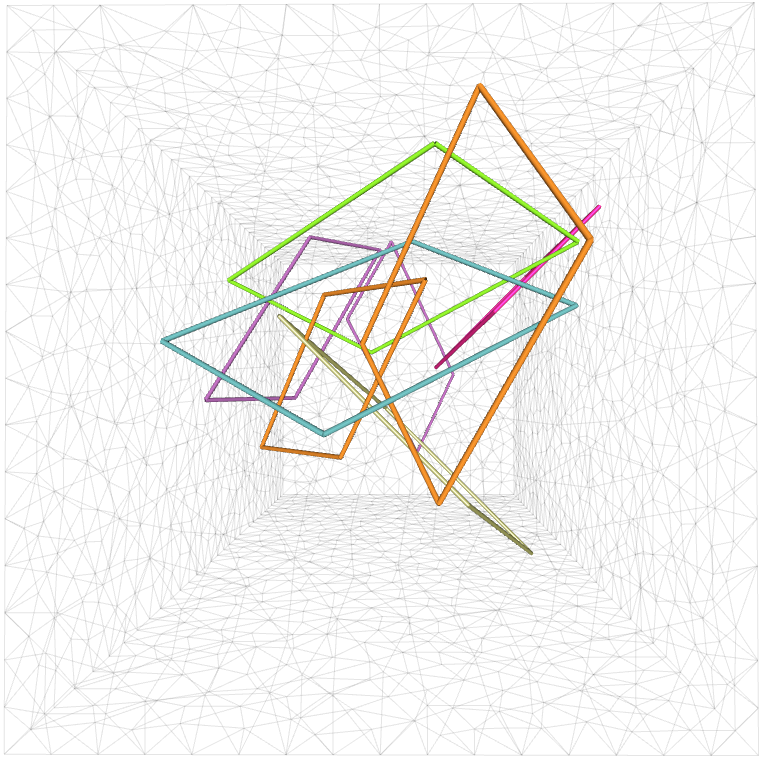}}
\subfigure[$\theta = 0.02$]
{\includegraphics[width=0.40\textwidth]{\figpath/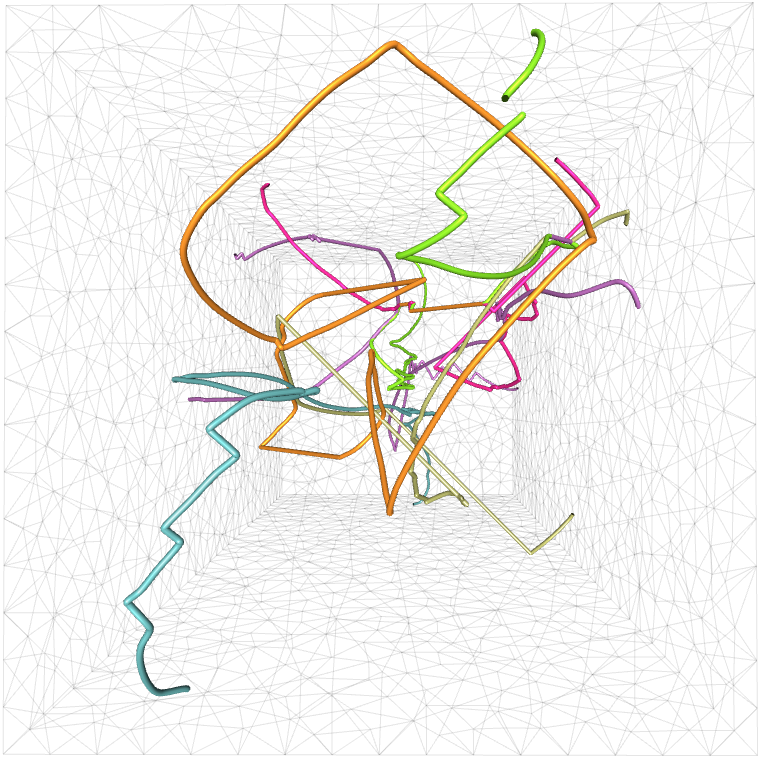}}
\subfigure[$\theta = 0.04$]
{\includegraphics[width=0.40\textwidth]{\figpath/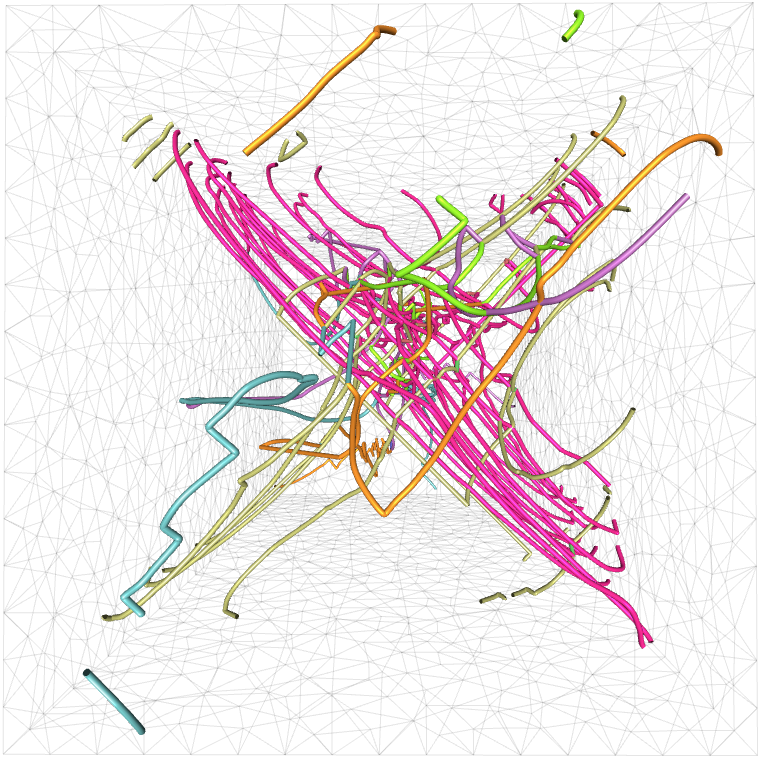}}
\subfigure[$\theta = 0.06$]
{\includegraphics[width=0.40\textwidth]{\figpath/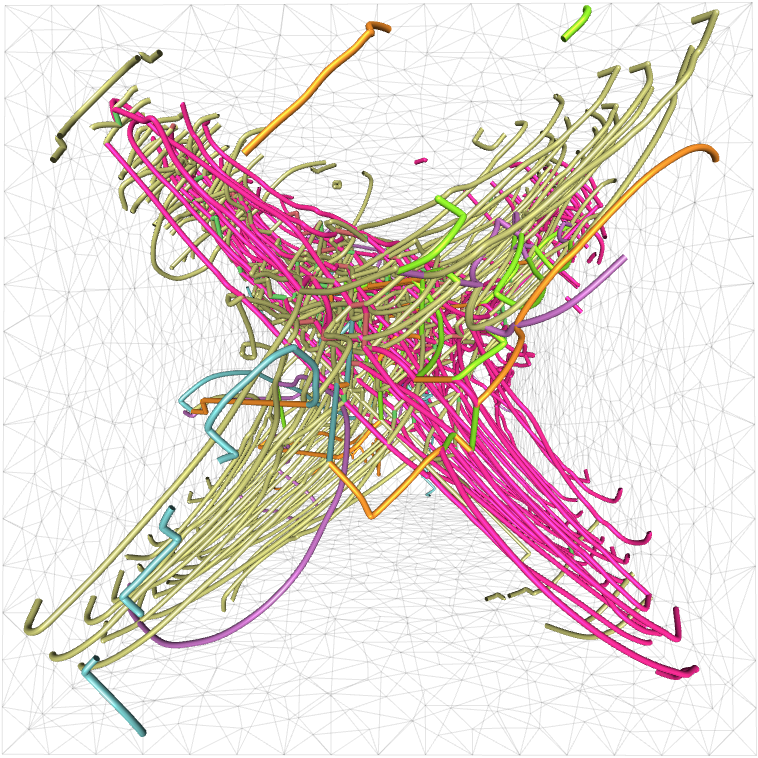}}
\caption{{Torsion with orientation $\langle 100 \rangle$: evolution of the dislocation network 
initially composed of a random distribution of glissile prismatic loops viewed along the twist axis.
Colors indicate dislocations lines with different Burgers vectors and the jogs in the dislocation lines are due to cross-slip events.
}}
\label{fig:torsion_evolution}
\end{figure}

The torque-twist response for both models are shown in \fref{fig:torsion_reaction}.
The correspondence between the calibrated CP model and the DD response for the $\langle 100 \rangle$ orientation of the wire  is similar to that shown in \fref{fig:shear_reaction_vs_strain} for shear. 
For the $\langle 110 \rangle$ oriented wire, however, the DD response shows a delayed onset of significant plasticity, as evidenced both by the torque-twist response as well as the line density evolution, whereas the CP torque-twist for this orientation is nearly the same as for $\langle 100 \rangle$.
\fref{fig:torsion_evolution} shows the evolution of the dislocation network with twist for the $\langle 100 \rangle$ case.
The dislocations are primarily screw dislocations on different glide planes.   
They reach their shown locations after a series of cross slip events from a few initial common sources.
Clearly, between twist $\theta = 0.02$ radians and $0.04$ radians, the initial random network multiplied through Frank-Read mechanisms (no nucleation model is employed) and aligned with the diagonals of the square cross-section. 
This regime also corresponds to the apparent onset of plasticity shown in \fref{fig:torsion_reaction}.

\begin{figure}[htbp]
\centering
\subfigure[ line density $\rho$]
{\includegraphics[width=0.38\textwidth]{\figpath/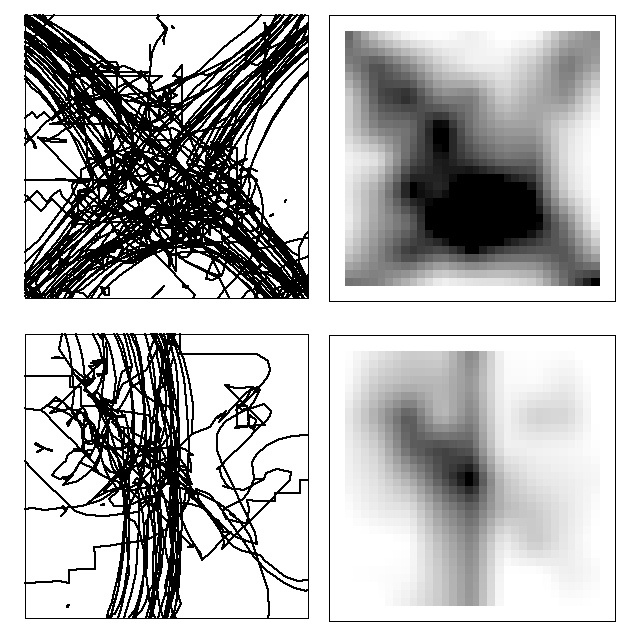}}
\subfigure[ resolved stress $\sb_i\cdot\stress \nb_i$]
{\includegraphics[width=0.58\textwidth]{\figpath/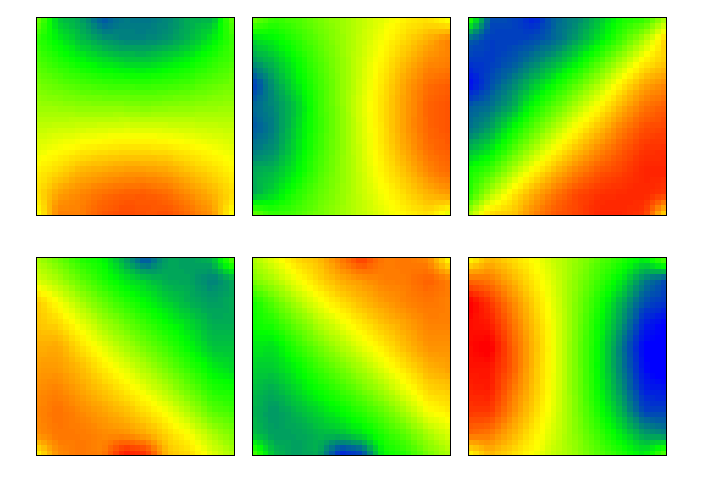}}
\caption{{Torsion:} (a) dislocation lines (left), line densities (grayscale) and (b) resolved shear stress $\sb\cdot\stress \nb$ on plane with normal $\nb=$(1,1,1) projected onto the $x_1$-$x_2$ plane for $\sb=$(1,1,1), (-1,1,0), and (1,-1,0) (rainbow).
The $\langle 100 \rangle$-orientation is shown on the top row and the $\langle 110 \rangle$-orientation on the bottom row.
}
\label{fig:torsion_shear_stress}
\end{figure}

\fref{fig:torsion_shear_stress}a shows the dislocation network and resulting line density at distinctly plastic state, twist angle $\theta$ = 0.06 radians for both the $\langle 100 \rangle$ and $\langle 110 \rangle$ cases.
In the $\langle 100 \rangle$ case, dislocation lines spanning the diagonals of the wire's square cross-section pile up; whereas, in the $\langle 110 \rangle$ case shown, the majority of the lines are vertical and span the opposing face mid-lines while a minority span one pair of the diagonal corners as in the $\langle 100 \rangle$ case.
Clearly from the line networks themselves and the resulting line density, the dislocation lines in the $\langle 110 \rangle$ case are more dispersed spatially and less numerous in total than in the $\langle 100 \rangle$ case.
(Note these networks are still dense relative to the nominally homogeneous cases where there is no tendency for dislocations to accumulate in any particular region of the body.)
In DD, dislocation lines are generated in high stress regions near the corners of the wire in torsion and migrate to the low/zero stress basins shown in green in \fref{fig:torsion_shear_stress}b, given the Peach-Kohler forces in \eref{eq:PMEP}.
For the two cases, the orientations of the slip planes relative to the wire are different, but elastic stress field is same, up until yield, due to the isotropic moduli assumption.
The resolved shear stress (RSS) is plotted in \fref{fig:torsion_shear_stress}b for both cases. 
In the $\langle 100 \rangle$ case the diagonal basin in RSS is dominant and in the $\langle 110 \rangle$ case the vertical basin is dominant but broader than in the $\langle 100 \rangle$. 

\begin{figure}[htbp]
\centering
\subfigure[\ DD ]
{\includegraphics[width=0.45\textwidth]{\figpath/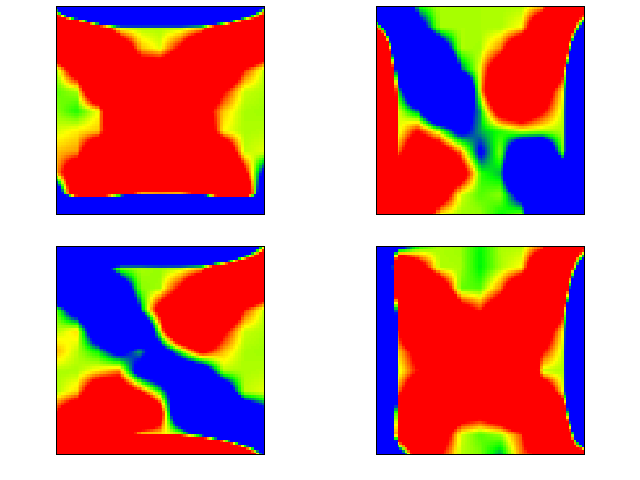}}
\subfigure[\ CP ]
{\includegraphics[width=0.45\textwidth]{\figpath/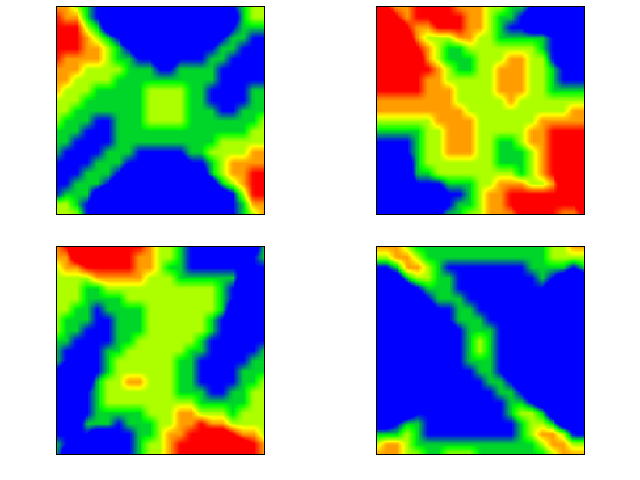}}
\caption{Torsion with orientation $\langle 100 \rangle$: comparison of
$\dislocationdensity_{11}$ (upper left),
$\dislocationdensity_{12}$ (upper right),
$\dislocationdensity_{21}$ (lower left),
$\dislocationdensity_{22}$ (lower right) resulting from (a) DD and (b) CP simulations at twist = 0.006 radians.
Here, red corresponds to $\dislocationdensity = 0.00258 / \ell$ and blue to $\dislocationdensity = -0.00258 / \ell$.
}
\label{fig:torsion100}
\end{figure}

\begin{figure}[htbp]
\centering
\subfigure[\ DD ]
{\includegraphics[width=0.45\textwidth]{\figpath/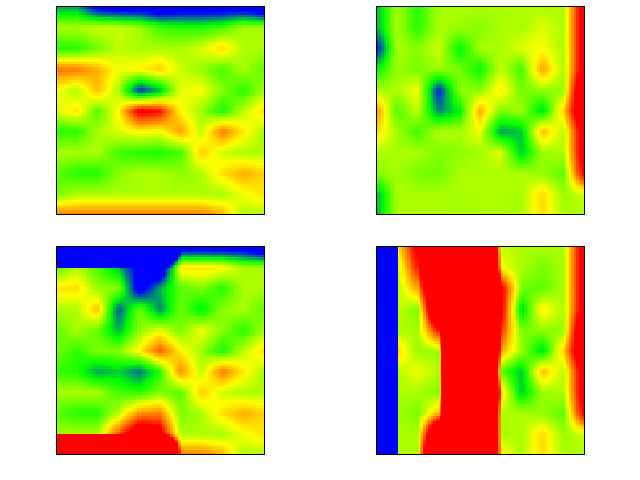}}
\subfigure[\ CP ]
{\includegraphics[width=0.45\textwidth]{\figpath/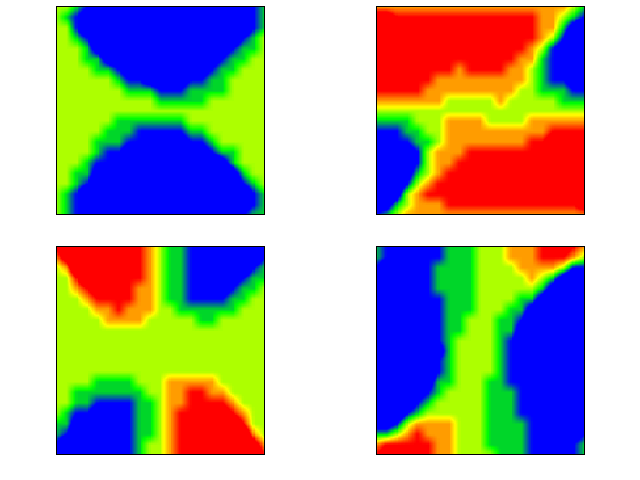}}
\caption{Torsion with orientation $\langle 110 \rangle$: comparison of
$\dislocationdensity_{11}$ (upper left),
$\dislocationdensity_{12}$ (upper right),
$\dislocationdensity_{21}$ (lower left),
$\dislocationdensity_{22}$ (lower right) resulting from (a) DD and (b) CP simulations at twist = 0.006 radians.
Here, red corresponds to $\dislocationdensity = 0.00258 / \ell$ and blue to $\dislocationdensity = -0.00258 / \ell$.
}
\label{fig:torsion110}
\end{figure}

Since our system size and total line density are at the threshold of being able to be resolved given the estimated minimum asymptotic scale, we average the DD data through the $x_3$-axis since slices of the solution are nominally equivalent along the twist axis.
\fref{fig:torsion100} and \fref{fig:torsion110} show the two dimensional projection of the in-plane components of the dislocation density tensor for the $\langle 100 \rangle$ and $\langle 110 \rangle$ orientations of the wire, respectively.
In general, the $\dislocationdensity$ fields generated with DD and CP are only comparable up to a divergence-free/solenoidal field \ie
$\Fb_p$  gradient of scalar function does not contribute to $\dislocationdensity$ (which follows directly from a Helmholtz decomposition of $\Fb_p$).
Also, the solutions have nominal 90 degree rotation and inversion symmetry.
The fields resulting from both DD and CP have comparable magnitudes.
And yet, the $\dislocationdensity$ fields generated by DD have significant components in the center of wire, unlike CP.
In the $\langle 100 \rangle$ orientation case, the magnitudes of $\dislocationdensity$ components are similar; but in the $\langle 110 \rangle$ case the weak pattern correspondence breaks down due to more dispersed nature of dislocation line density.
As mentioned, in DD lines are generated at high stress regions at the lateral edges of the prismatic wire and migrate to the RSS basins either diagonal or crossing the midpoints of the wire's cross-section.
In contrast, in a CP simulation, a background of mobile, geometrically necessary dislocations are assumed to be present everywhere and plastic slip occurs in regions of high stress. 
In CP there is no mechanism for explicit generation of dislocations nor compatibility of plastic slip induced by the motion of the implicit dislocation network.
Hence, the dislocation density field produced by CP only responds to the regions of high RSS, and there is no driving force for migration/flow of dislocation density to low RSS basins.

\begin{figure}[htbp]
\centering
\subfigure[ $m$=1]
{\includegraphics[width=0.300\textwidth]{\figpath/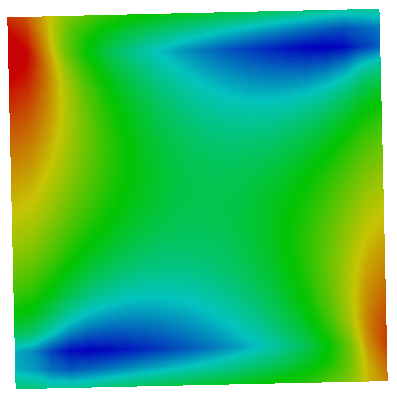}}
\subfigure[ $m$=4]
{\includegraphics[width=0.315\textwidth]{\figpath/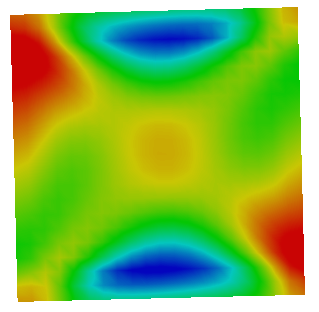}}
\caption{{Torsion:} dislocation density $\dislocationdensity_{11}$ for flow rule exponent $m$=1,4.
}
\label{fig:torsion_m_comp}
\end{figure}

As a last comparison, \fref{fig:torsion_m_comp} shows the sensitivity of the dislocation density field for the $\langle 100 \rangle$ case predicted by CP to the flow rule \eqref{eq:flow_rule} exponent $m$. 
Clearly the significant dislocation density approaches the center of the cross-section and the ``$\times$'' pattern becomes more prominent and sharper with increased $m$ showing better correspondence with the DD results.
An explanation for this better correspondence is discussed in the concluding section.
Due to numerical issues with the form the simple flow rule \eqref{eq:flow_rule} it was not possible to increase $m$ to further improve the correspondence with the DD results.

\section{Discussion}\label{sec:dis}

In this work, we have established that a simple CP model calibrated to DD data can produce similar microstructural fields as well as overall response.
This similarity is predicated on resolving the dislocation density field at a scale above the intermediate asymptotic length scale intrinsic to the material with its native dislocation population.
For the material we examined, Cu with line density $\rho = 10^{13}$ m$^{-2}$, we established that the minimum asymptotic scale is approximately 500 $b$.
Physically, the asymptotic scale is determined by how densely dislocation lines pack after being generated at high RSS regions and migrating to low RSS regions. 
In particular, we estimated the scale at which continuum dislocation density fields should appear with nominally homogeneous simulations where there is little difference between high and low RSS regions.
This estimate proved robust in constructing dislocation density fields in the inhomogeneous loading cases we considered.
For the finite systems we examined, it appears that the dislocation pile-ups on particular slip planes reach a characteristic spacing due to the fact that forces between dislocations scale as $b^2$, while forces due to the applied stress scale as $b$
(refer to Chap. 21 in \cite{Hirth:1992ub}).%
\footnote{Recall $b$ is the Burgers vector of the material.}
Hence, we conjecture that the asymptotic scale for the dislocation density is strongly dependent on material and less dependent on the particular loading case.

By examining \eref{eq:hardy_alpha} and \eref{eq:hardy_rho}, the connection between the dislocation line density $\rho$ and dislocation density tensor $\dislocationdensity$ can be seen.
Clearly $\rho$ indicates the general magnitude of $\dislocationdensity$, \ie there are no $\dislocationdensity$ contributions where $\rho$ is zero; but, $\dislocationdensity$ adds vectorial information from the line direction as well as from the Burgers vector, see \eref{eq:hardy_alpha}
In contrast, with a CP model the dislocation density $\dislocationdensity$ results from the curl of the plastic deformation gradient through the CP flow rule, \eref{eq:flow_rule}.
So, unlike the explicit representation of dislocation lines in DD, 
CP, through its flow rule, conflates generation and migration and only creates mobile slip in high stress/deformation gradient regions.
Moreover it simulates the hardening due to entanglement of dislocation lines with the flow rule.

A loose correspondence can be made between the CP flow rule and the dynamics of the dislocation lines that governs DD evolution. 
The Peach-Kohler forces driving the dislocation dynamics are analogous to the resolved shear stresses driving plastic evolution in the flow rule so that the dislocation line velocity is proportional to the Peach-Kohler force, and the plastic slip rate is proportional to the local resolved shear stress. 
From our results, is it clear that, although the global response of corresponding DD and CP models can be made the same, the microstructural dislocation density tensor field is much more sensitive to details of the calibrated flow rule.
Although algorithmic stability issues prevented us from exploring the response of flow rules with high exponents, we were able to show that increased exponent tends to improve the qualitative similarity of the in-plane $\dislocationdensity$ fields for a prismatic wire in torsion where there is a strong tendency for the dislocation network to align with geometric features.
It stands to reason that this improvement will saturate before the $m\to\infty$ limit is reached, where plastic slip, and by analogy dislocation velocity, is insensitive to RSS magnitude.
In fact, it is hard to imagine how a flow rule could be constructed to perfectly represent the $\dislocationdensity$ patterns seen with DD unless some compatibility of the plastic slip representing dislocation motion is enforced. 
This modeling direction is currently being developed within the framework of \textit{continuum dislocation theory}, as in \cite{Acharya:2006uz}, \cite{Xia:2015dh} and \cite{Hochrainer:2014kf}, where dislocation densities  are independent field quantities evolving according to transport equations, and plastic slip is computed from dislocation fluxes.

In general, the continuum field, $\dislocationdensity$ given by Eqs. \eqref{eq:dd_macro_reduce} and \eqref{eq:hardy_alpha}, captures the (geometrically necessary) dislocation density and is a quantitative measure useful in understanding the microstructures that result from large strains in the plastic regime. 
This tensor can be used in the yield function of macroscopic phenomenological theories of plasticity to reproduce realistic dislocation microstructures. 
Such models form alternatives to the crystal-plasticity formulations based on active slip systems. 
For example, \cite{Edmiston2013} include the DDT in the yield function, and develop a rate-independent macroscopic theory based on material symmetry. 
As the authors demonstrate, inclusion of DDT in the yield function introduces an additional length-scale into the macroscopic theory. 
Using this length-scale as a parameter, \cite{Edmiston2013} have shown the formation of dislocation microstructures with spatially localized DDT similar to our work. 
For similar structures, the value of the length-scale in the theory are also similar to the range of intermediate asymptotic length scales that we estimated are needed to obtain a field description of the DDT from the DD simulations. 
This suggests that performing DD simulations to evaluate the DDT tensor may serve as a practical tool for calibrating and validating the phenomenological theories of plasticity. 
Also the correspondence between DD and CP could perhaps be improved  by using a flow rule with explicit dependence on the DDT computed from the plastic deformation gradient.
An analysis along these lines is left for future work.

\section*{Acknowledgements}
The authors would like to acknowledge helpful discussions with J. Ostien, A. Mota, F. Abdeljawad, H. Lim, and R. Sills (Sandia), and support from Sandia National Laboratories' Advanced Simulation and Computing/Physics \& Engineering Models (ASC/P\&EM) program.
Sandia National Laboratories is a multi-program laboratory managed and operated by Sandia Corporation, a wholly owned subsidiary of Lockheed Martin Corporation, for the U.S. Department of Energy's National Nuclear Security Administration under contract DE-AC04-94AL85000. 

%
%
\bibliographystyle{unsrt}

\end{document}